# Using Virtual Reality in Museums to Bridge the Gap Between Material Heritage and the Interpretation of its Immaterial Context

Carlos R. Cunha[1][0000-0003-3085-1562], Vítor Mendonça[2][0000-0001-7020-8235], André Moreira[2][0000-0002-6253-6615], João Pedro Gomes[2][0000-0001-9308-0027] and Aida Carvalho[3][0000-0001-8997-9195]

[1] UNIAG, Instituto Politécnico de Bragança, Campus de Santa Apolónia, 5300-253 Bragança, Portugal
`crc@ipb.pt`

[2] Instituto Politécnico de Bragança, Campus de Santa Apolónia, 5300-253, Bragança, Portugal
`mendonca@ipb.pt, andre-moreira@ipb.pt, jpgomes@ipb.pt`

[3] CiTUR, Instituto Politécnico de Bragança, Campus de Santa Apolónia, 5300-253 Bragança, Portugal
`aidacarvalho@arte-coa.pt`

**Abstract.** Material heritage typically has a whole set of associated immaterial heritage, which is essential to pass on to the visitor as a cultural mission of the destinations and those who manage them. In this sense, the interpretation of material heritage is a complex process that is not a fully efficient process with the mere observation of physical artefacts. In this context, it emerges as fundamental to provide visitors with a set of tools that allow them to correctly interpret the artefacts that come to fully understand the cultural dimension of the destinations and their heritage. Accordingly, the role of Virtual Reality can leverage the creation of innovative and immersive solutions that allow the visitor to understand and feel part of their own heritage and its ancestral component that defines the sociocultural roots of destinations and their civilizational traditions. This article, after dissecting and substantiating the role of Virtual Reality in the interpretation of heritage, presents a conceptual model, based on the use of Virtual Reality, which was, in part, prototyped in the scenario of the Portuguese Museum in the city of Miranda do Douro. This proposal is an ongoing contribution to the creation of innovative and immersive tools for the interpretation of heritage.

**Keywords:** Virtual Reality, Museum, Heritage, Interpretation., Model, Prototype.

## 1. Introduction

According to [1], human civilization translates a set of material and immaterial conquests in which none of them was produced without the other. In this context, it is also mentioned that the intangible cultural heritage is part of the identity of civilizations. The cultural heritage of nations is composed of material and immaterial heritage. In this context, [2] refers to the recognition of the intangible layer of cultural heritage, established by the UNESCO Convention of 2003 that highlights the Safeguarding of Intangible Cultural Heritage.

Not infrequently, material heritage has associated with it a vast and rich set of intangible cultural heritage that must be presented and explained to the visitor so that he can understand the full richness of the physical objects he sees. According to [3], the ability to enable the interpretation of destinations is important to help tourists obtain the authentic values of destinations-heritage and, in this way, promote an effective interpretation of tourism in the delivery of natural and cultural values – "perception of heritage values is influenced by interpretive information and interpretive principles".

Museums, as well as many of the visitable collections, are full of artifacts that have an enormous cultural richness that is not always easily known to the public, which complicates the process of interpretation. This difficulty is increased when the visitor is a foreigner and knows little or almost nothing about the culture of the country he visits. In this context, the use of tools that allow the visitor a greater ability to interpret the objects he sees is essential and will make all the difference in the fruition of the visit. For respond to this challenge, technology has been embedded in museums to provide more information to the visitor. In this domain, Virtual Reality (VR) has been widely considered a technology capable of allowing visitors to obtain information, in an immersive approach, about the museum's collections [4].

Technologies capable of creating immersive environments, such as VR, are tools that can play a very important role in attracting new visitors, due to their ability to provide incredible experiences and maximize the likelihood of the visitor to repeat and recommend the visiting experience [5]. A vision of a digital museum requires an analysis of how museums can take advantage of technology to boost the development of innovation networks, competitive advantages and leverage their economic performance; as well, as solutions for digitizing and improving the visitor experience [6]. The development of VR solutions must be based on the ability to allow users greater involvement, interactivity and control over their experiences; this will positively influence the emotional response and the intention to spread word of mouth about the VR experience itself [7].

Emerging digital technologies are a critical success factor for the sustainable preservation and communication of cultural heritage to the general public [8]. In this context, museums are no exception, as they should be able to innovate in the way they communicate to visitors the cultural heritage that they contain in their exhibition space. Providing mechanisms that enable visitors to use effective forms of interpretation is therefore central to the success of any museum's mission.

This article begins by explaining the correlation between the material heritage and its immaterial context, and the importance of making known, in an effective and motivating way, the entire cultural context of the artefacts, to those who visit a museum. Then, a review of the state of the art is made on the potential of VR as a link between the physical and the virtual, and as an enhancer of the visitor's ability to interpret. After this framework, a conceptual model capable of adapting to the museological context and the technological integration with its objects is proposed, incorporating the potentialities of VR. Finally, an experimental prototype, developed for the Museum of the Portuguese city of Miranda do Douro, is presented, where most of the exposed physical objects are linked to ancestral traditions (in which these objects were used), in an approach capable of improving the interpretive effectiveness of the visitors and their ability to understand the total cultural dimension of the objects exhibited in the museum.

## 2. Material Heritage and its Immaterial Context

The 1999 ICOMOS International Charter on Cultural Tourism states that "(...) the specific heritage and collective memory of each community and each place are irreplaceable and represent an essential basis for a development which is both respectful of the past and forward-looking. (...) in this age of increasing globalization, the protection, conservation, interpretation and dissemination of the heritage and cultural diversity of each place or region constitute a major challenge for all peoples and all nations."

In this sense, there is a concern with the forces of globalization and cultural homogenization and the search for cultural identity. The concept of Heritage is subject to numerous interpretations, influenced by changes in society itself and which have resulted in an increasing complexity.

It is the cultural legacy that we have received from the past, that we live in the present and that we will pass on to future generations (Convención de 1972 para la Protección del Patrimonio Mundial Cultural y Natural la UNESCO).

From a conceptual point of view, cultural heritage is today the most comprehensive concept, integrating material heritage - which includes immovable and movable heritage - and immaterial heritage. In turn, mobile heritage comprises the set of works of art (e.g. painting, sculpture, furniture) and the immovable heritage comprises the architectural, archaeological and landscape heritage. In Portugal, the legal reference framework for the whole universe of cultural heritage is given by the Law for the Protection and Enhancement of the Cultural Heritage - Law 107/2001, of 8 September. In this framework, the concept of patrimonial heritage is constituted by the universe of material and immaterial assets of relevant cultural interest, as well as "the respective contexts which, due to their value as witnesses, have an interpretative and informative relationship with them", Relevant cultural interest is understood as its "historical, paleontological, archaeological, architectural, linguistic, documental, artistic, ethnographic, scientific, social, industrial or technical" value, reflecting "values of memory, antiquity, authenticity, originality, rarity, singularity or exemplarity".

Tourism enables the use of heritage and cultural resources are one of the drivers of tourism growth. However, the mere fact that heritage resources exist does not imply the existence of tourism resources or attractions. It is fundamental to structure the offer through the creation of diffusion, reception, and interpretation centers, implementing processes of structuring and organization of heritage assets for their cultural tourism use. There is an ecosystem of value that needs to be created, i.e. heritage resources need means of interpretation for the attribution of meaning and value to become a cultural resource; and, in turn, this cultural resource needs means of dissemination and promotion to enter the cultural tourism offer.

In order to the visitor to have a full interpretive experience of heritage and destinations, the creation of interpretation mechanisms based on Virtual Reality may constitute a leveraging approach, due to the ability to create immersive and innovative experiences that involved the visitor and they imbued it in the atmosphere with the culture of material heritage and all its valuable immaterial component.

## 3. Virtual Reality and the Heritage Interpretation

According to [9], the interpretation of heritage is considered an effective tool for learning, communication and management, capable of increasing the visitor's awareness and empathy for the heritage sites or artifacts he visits or contacts. [10], presents an empirical proposal focused on the importance of using modern devices and technologies in heritage interpretation and tourism marketing. In this context, it is important to develop innovative strategies that enable the best possible interpretation. The use of immersive technology, such as VR, has enormous potential to create memorable tourist experiences, specifically in heritage tourism [11].

There is an exponential growth of cultural heritage across the "digital ecosphere" which has caught up with us, regardless of traditional factors of space or time [12]. In the domain of VR, VR applications are able to show virtual reconstructions of a certain cultural property and allow and improve the user's understanding and interpretation of that heritage [13]. According to [14], the incorporation of 3D artifacts into VR experiences has been of interest to Cultural Heritage professionals. The use of VR can support on-the-spot interpretation experiences – translating physical objects into their non-physical cultural context; as well, it can be used as a form of remote and online interpretation. Within its remote use, according to [15], 360° VR interpretation and presentation technology can help many tourists remotely learn the enormous value of the world's cultural heritage.

If, on the one hand, the use of immersive technologies has been applied in various scenarios in the context of tourism and heritage, we think that the creation of conceptual models capable of capitalizing on the interpretive experience of the visitor, equally, in favor of the manager of the cultural space, is something which still needs more research and contributions. A visit must be an act of interpretation and satisfaction for the visitor, at the same time it must be able to provide information that allows managers and curators of cultural spaces a better capacity to manage these spaces.

In our opinion, another factor that must be guaranteed to the visitor is the ability to interpret the space, in fact. With the globalization of tourism, the concept of visitor is now multicultural and with a huge geographic dispersion of origin. Thus, it is important to meet the measure of knowledge and sensitivity of each visitor, if we want the interpretation to be effective and the cultural message to be effectively understood. According to [16], the use of technology in the context of an exhibition can allow the collection of diverse demographic data and increase the reach of a museum, such as presenting the interpretation of digital heritage in an innovative way.

The next chapter proposes a conceptual model capable of using VR to provide experiences of heritage interpretation, linking cultural objects to their rich immaterial context, while this VR experience generates useful data for the space manager.

## 4. Proposed Conceptual Model

Whereas virtual museums, also, add new possibilities to the exhibits, allowing for a new relationship between the object and the visitor in ways that physical museums do not provide [17], and allowing the visitor to enjoy immersive and complementary experiences, we believe that the development of VR museums applications should allow users to interact with different elements and contents, and this diversification will enrich the experience of visiting museums.

With the purpose of developing a virtual reality application, which, in our opinion, should be based on dynamic content to diversify the experience of visitors. As such, it will be up to the museum manager to feed a repository of different contents (e.g. images, videos, 3D objects), as well as to define the interaction elements (Assets) that will trigger the visualization of these contents.

In this context, and according to the proposed model, figure 1, the VR Engine establishes the intermediation between the visitor and the contents to be made available due to the interactions carried out and the contents available for viewing. Therefore, the VR Engine will have to include a set of rules whose responsibility will be the display of contents stored in the museum's internal repository (Internal Data Base) or in repositories external to the museum (External Data Bases), that is, existing content from other entities.

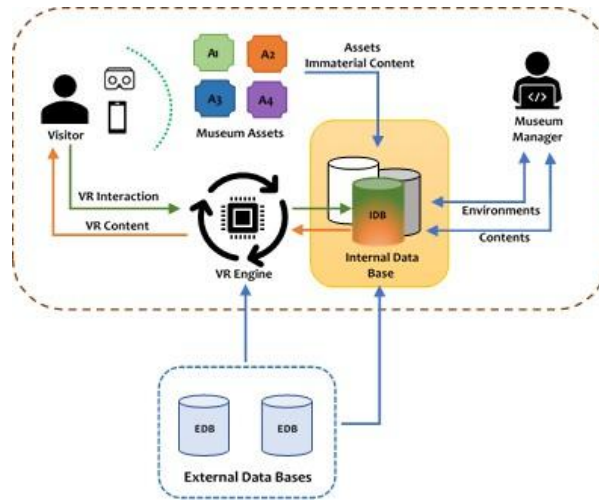

**Fig 1.** Proposed Conceptual Model

Access to existing content in external repositories aims to provide the model with greater elasticity, considering that around the museum there may be various content produced and managed by other entities (e.g. City Council, Universities, Cultural Associations) that may be relevant and complementary to the contents managed by the museum. For example, the museum manager could map a particular Asset to display a video produced by a folklore group.

Considering that the content is intended to be dynamic, we should point out that, the museum manager can whenever he deems it appropriate to add or remove Assets and/or Contents.

To implement the proposed model, the proposed system includes a back office and a front-end application. Figure 2 summarizes the features that the system makes available to the museum manager, that is, the back office features.

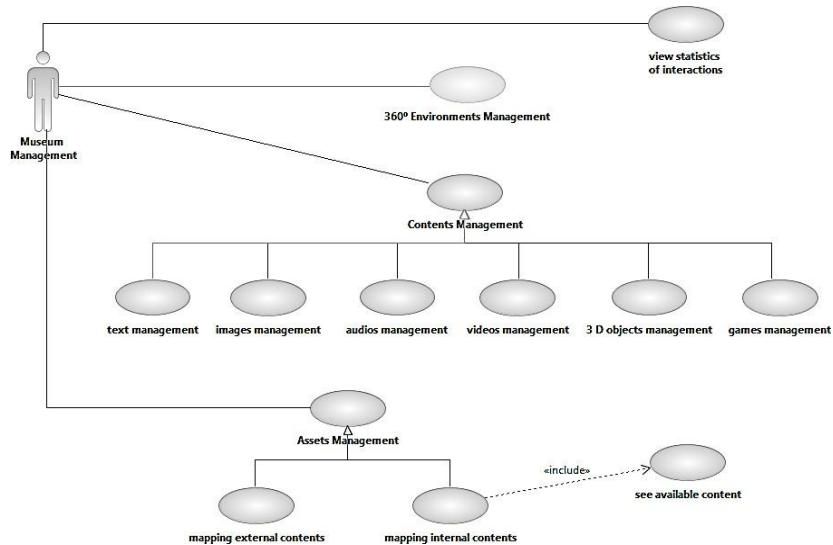

**Fig 2.** Use Cases Diagram: Management Features

The front-end with the visitor (user) is made using Virtual Reality visualization and interaction equipment, especially with virtual reality glasses. The use case diagram shown in Figure 3 summarizes the features that the application makes available to the user for a virtual reality experience.

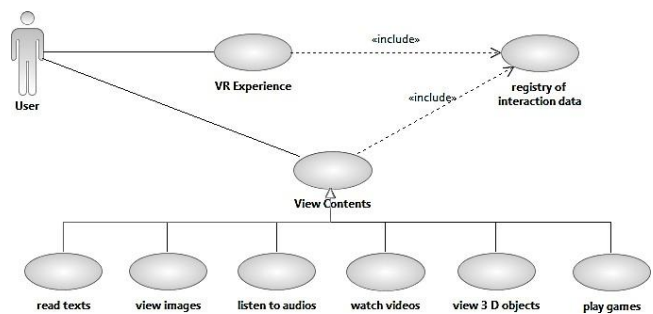

**Fig 3.** Use Cases Diagram: Visitor Experience

Thus, and according to the diagrams presented above, the museum manager will be able to manage the 360º virtual environments, manage the contents (text, images, audio, videos, 3D objects and games) and map the Assets that will allow triggering the interaction with the different contents, in turn, the visitor (user) will be able to enjoy the VR experience and when interacting with the Assets will have access to the different linked contents, allowing the visitor to complement their experience, that is, being able to read texts, view images, listen to audios, visualize videos, interacting with 3D objects, or playing a game.

The use register of user interactions will allow the museum manager to consult the statistics of these interactions to make decisions about the type of content to be made available.

## 5.  An Experimental Prototype Applied to the Museum of Miranda do Douro City

To carry out experiments with the use of VR to link the immaterial context of the "Terra de Miranda" museum assets exhibition, an experimental prototype was developed, which allowed to corroborate the model presented above.

Thus, the prototype objectives, in addition to allowing the model to be tested, also intend to demonstrate that the combination of different immersive virtual contents (text, image, audio, video, 3D objects and/or games) will enrich the visitor's experience.

As for the technical specifications of the development, the prototype that was developed have used the VR equipment HTC® VIVE® Pro EYE full kit (figure 4), running on a computer with an Intel® Core™ i7-10700KF processor with 32 GB of RAM and a NVIDIA® GeForce® RTX 2070 SUPER ™ graphics card. The computer features exceed, by far, the minimum requirements of VR equipment and guarantee an excellent performance of the VR experience.

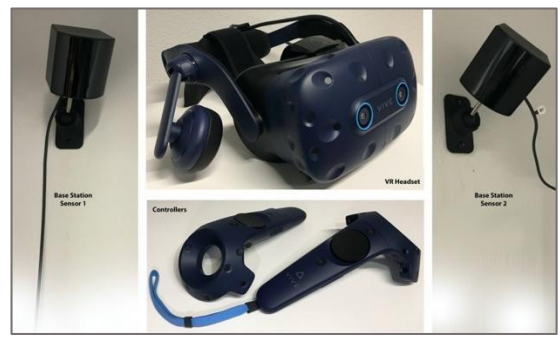

**Fig 4.** VR Equipment – HTC® VIVE® Pro EYE full kit

The development process started with the capture of 360º images of the different spaces of the museum (figure 5). For the creation of 360º images, a Xiaomi Mi Sphere camera was used.

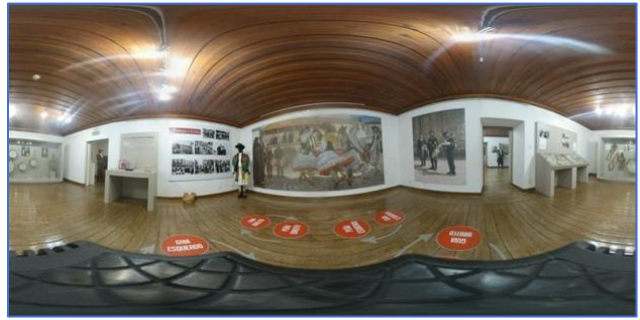

**Fig 5.** Example of a 360º image museum-space environment

The programming of the experimental scenario and integration of assets, the Unity game engine was used along with the C# programming language. In this way, the areas and objects on display were mapped. An example is illustrated in figure 6. In parallel to this phase, the multimedia contents to be indexed to each object in the museum were compiled and the dynamics of interaction between the user and the different objects designed.

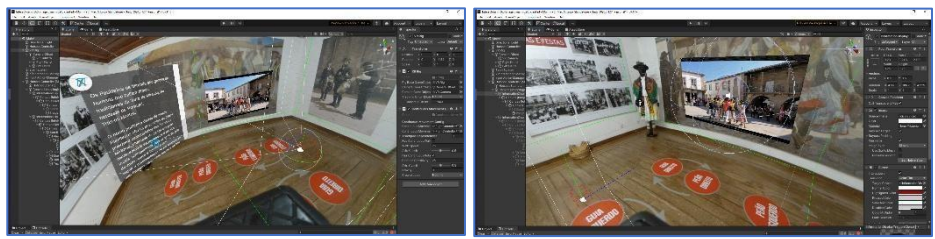

**Fig 6.** VR assets and information-linking mapping-process

After surveying the objects exhibited in the museum and the digital contents that they wanted to associate, the inclusion of marker Assets was carried out to allow the visitor to identify which objects have associated digital contents.

In this context, Assets were mapped and associated with digital content, as well as the graphic way in which they will be displayed (e.g. pop-up with text, video). Said contents may come from the museum's internal repositories or even from external entities' repositories.

Accordingly, as illustrated in Figure 6, assets were mapped connecting the "Pauliteiro" mannequin to a descriptive text content, as well as the painting of traditional dances displayed on the panel to a video demonstrating the performance of unique traditional dances existing in the region of Miranda do Douro.

This entire process merges into a VR application that supports an immersive visit experience rich in diverse immaterial cultural content.

The experimentation of the developed application, illustrated in Figure 7, shows the example of the visitor's interaction with the different assets embedded in the virtual environment.

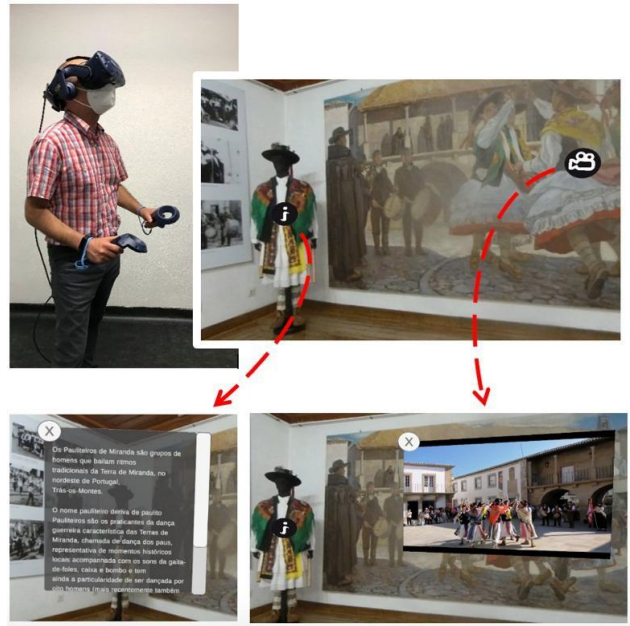

**Fig 7.** VR visitor experience scenario

Thus, when interacting with the existing Asset on the "Pauliteiro" mannequin, the visitor will be able to read a descriptive text of its characterization, history and about the traditional costume. As for the interaction with the Asset displayed on the panel, the visitor can watch a video (external source: https://youtu.be/iF6BUQ5sh-k).

Simultaneously, the application collects data about interactions between the visitor and the assets, as well as the visualization of the associated contents.

## 6. Conclusion

The material and immaterial heritage of regions reflects the culture and customs of the peoples, and museums are one of the best examples of a collection of excellence, whose mission is to preserve the cultural legacy between generations. However, a visit to a museum must guarantee the visitor a full interpretation experience.

In this context, the introduction of technology capable of providing innovative and immersive interpretation experiences is fundamental, especially given the degree of demand of the current (and future) visitor.

Accordingly, VR (along with other technologies) has the ability to create environments rich in digital content linked to physical objects and, in this way, translates into an approach capable of complementing and enriching the museum visitor's experience, maximizing the interpretive capacity of the exposed heritage.

This paper presents a VR-based model capable of providing an immersive experience to the visitor, providing them with useful digital elements for the interpretation of the museum's space, while ensuring useful feedback for the museum's managing entity, as well as the management of contents.

Based on the model presented, an experimental prototype was developed, which, at this stage, implements the VR component as a tool for mapping digital content, associated with the collection of the Miranda do Douro museum - "Museu da Terra de Miranda".

The work presented is part of a long-term effort to provide innovative approaches in tourism related industry.

## Acknowledgments

Anonymous funding for review purpose.